\documentclass{aa}
\usepackage{amsmath,amsfonts,amssymb}
\usepackage{textcomp,gensymb}
\usepackage[T1]{fontenc}
\usepackage{aecompl}
\usepackage{ulem}
\usepackage{graphicx}
\usepackage[usenames]{xcolor}
\usepackage{color}
\usepackage{times}
\usepackage{subfig}
\usepackage{stfloats}
\usepackage[utf8]{inputenc}
\usepackage{hyperref}
\usepackage{multirow}
\usepackage{enumerate,tabularx}
\usepackage{physics}
\usepackage{chemarrow}
\usepackage{tikz}
\usepackage{txfonts}
\usepackage{booktabs}
\usepackage{tablefootnote}
\usepackage{adjustbox}
\usepackage{mathtools}
\usepackage[flushleft]{threeparttable}
\bibpunct{(}{)}{;}{a}{}{,}

\usetikzlibrary{shapes,arrows}
\usetikzlibrary{shapes.geometric}

\newcommand{\rev}[1]{\textcolor{black}{#1}}

\begin{document}

    \title{Environmental interactions in Class II systems and their impact on the disk-planet architecture}

   \subtitle{}

   \author{Pedro P. Poblete\inst{1},
          Nicolás Cuello\inst{1},
          Tim D. Pearce\inst{2},
          Antoine Alaguero\inst{1},
          Josh Calcino\inst{3},
          Daniel J. Price\inst{4,1},
          \and
          Michael Kuffmeier\inst{5}.
          }

   \institute{Univ. Grenoble Alpes, CNRS, IPAG, 38000 Grenoble, France;\\
              \email{pedro.poblete@univ-grenoble-alpes.fr};
         \and
            Department of Physics, University of Warwick, Gibbet Hill Road, Coventry CV4 7AL, UK;
         \and
         Department of Astronomy, Tsinghua University, 30 Shuangqing Rd, 100084 Beijing, China;
         \and
             School of Physics and Astronomy, Monash University, Vic. 3800, Australia;
         \and
            Niels Bohr Institute, University of Copenhagen, Jagtvej 155a, DK-2200 Copenhagen, Denmark.
             }

   \date{Received September XX, XXXX; accepted March XX, XXXX}

  \abstract
   {Protoplanetary disks evolve in clustered environments where interactions with nearby stars and interstellar gas are common. Such environmental processes, including stellar flybys and gas infall, can significantly perturb disk structures over the disk lifetime and potentially influence the evolution of embedded planets. 
   }
   {We investigate how environmental interactions affect the architecture of Class II systems that host both a disk and already-formed planets, and assess their impact on disk structure and dynamics, as well as planetary evolution.
   }
   {We performed three-dimensional simulations using the {\sc Phantom} smoothed particle hydrodynamics (SPH) code, including multiple dust species treated with a dust-as-particles approach that accounts for dust back-reaction on the gas. We modeled a disk hosting two planets in a 2:1 mean-motion resonance and subjected the system to two types of perturbations: an infalling gaseous cloudlet and a stellar flyby. 
   }
   {Infall and flyby perturbations change the disk morphology and dynamical state. Infalling gas increases the disk mass, angular momentum, and dynamically excites the dust to produce eccentric and multi-ring dust structures. The stellar flyby truncates the disk, compacting the dust distribution radially and enhancing episodic radial migration of dust grains. These processes excite eccentricity in both gas and dust, leading to distinct accretion pathways for the planets. In particular, the flyby promotes inward dust migration that may enhance solid accretion by the planets, while infall preferentially increases the accretion rate of the inner planet.
   }
   {Environmental interactions during the Class II phase can reshape disk–planet systems, imprinting dynamical signatures that may persist into later evolutionary stages. Both late infall and stellar flybys influence the growth and composition of planets; in particular, infall events can lead to the formation of eccentric, narrow debris disks.
   }
  
{}

   
   \titlerunning{Environmental interactions in Class II systems}
   \authorrunning{Pedro P. Poblete et al.}
   \maketitle 

\label{firstpage}

\section{Introduction}
\label{sec:intro}

Protoplanetary disks (PPDs) are circumstellar structures that form naturally during the early stages of star formation. They are composed primarily of gas, with a smaller fraction of dust. These disks are where planets form, by the growth of dust grains into planetesimals and fully formed planets \citep[see][for a brief review]{Armitage2018}. The classical picture of PPD evolution assumes an isolated system, in which the star, disk, and forming planets evolve without external influences. While this framework is theoretically well motivated, it does not fully capture the environments in which disks are observed, as a substantial fraction of disks --- approximately 55\% in the Taurus star-forming region, for example --- are subject to interactions with nearby stars and infalling gas in the interstellar medium \citep{Garufi+2024}.

\rev{\citet{Pfalzner2013} defined a {flyby} as a single close encounter between two stars with a minimum separation smaller than 1000 au. The impact of a flyby on the perturbed system depends on the encounter strength, which \citet{Winter+2024} quantified in terms of the fraction of disk mass removed by the passing star. Encounters that remove $\sim 10\%$ and $\sim 1\%$ of the disk mass are classified as strong and weak, respectively. Based on this definition, the corresponding occurrence rates range from $\sim 10\ {\rm Myr^{-1}}$ for strong encounters to $\sim 200\ {\rm Myr^{-1}}$ for weak encounters.} When a star hosting a PPD undergoes a stellar flyby, the disk morphology and dynamics can be significantly altered through tidal interactions, excitation of eccentricity, and other perturbations \citep{Clarke&Pringle1993, Cuello+2019, Cuello+2023}. The consequences of flybys are not limited to the disk itself but also affect the architecture of embedded planets and protoplanets. \citet{Laughlin&Adams1998} demonstrated that flybys can induce orbital disruption and eccentricity growth in Jovian planets, \citet{Malmberg+2011} investigated planet ejection following stellar encounters, and \citet{Charalambous+2025} examined the disruption of planetary mean-motion resonance (MMR) chains.

Alternatively, star–disk systems can be influenced by the accretion of gas from the interstellar medium. {Infall} refers to the process by which a system captures and accretes molecular cloud gas. Such events are especially common during the early stages of planet formation ($\lesssim 1$ Myr) and are associated with the high accretion rates in the observed in young stars, occurring in more than 50\% of Class 0/I objects \citep{Dunham+2014,Hartmann+2016} and related to the Bondi–Hoyle–Lyttleton accretion process during the post-collapse phase \citep{Padoan+2005,Winter+2024c,Padoan+2025}. The occurrence rate depends on age, and in more evolved systems, namely Class II objects --- stellar systems dominated by a PPD and potentially hosting already formed planets --- the rate is estimated to be $\sim 15\%$ based on the census of the Taurus star-forming region \citep{Garufi+2024}, and $\sim 20\%$ of the total near-infrared sources observed to date \citep{Garufi+2026}. The primary consequence of infall for protoplanetary disks is their rejuvenation, as the supply of fresh gas replenishes the system and enhances the accretion rate onto the central star. This process may trigger FU Orionis–like outbursts \citep{Padoan+2014} and can potentially lead to a reclassification of the system toward earlier evolutionary stages, such as from Class I to Class 0 \citep{Kurosawa+2004,Dullemond+2019, Kuffmeier+2020, Kuffmeier+2023}. Additionally, infall can modify the disk’s angular momentum, potentially inducing misalignments and warps \citep{Bonnell1994,Bate+2003,Thies+2011, Wijnen+2017, Kuffmeier+2021,Kuffmeier+2024, Pelkonen+2025}. Despite these potentially important effects, the impact of infall events on planetary evolution remains poorly understood. Most existing studies focus on planetesimal formation through streaming instabilities and self-gravity \citep{Schib+2023,Huhn+2025,Longarini+2025}, and only speculative scenarios about how planetary migration might occur afterward \citep{Zhu+2010}.

The studies on flybys and infall discussed above have primarily focused on either the earliest stages of planet formation, when only a PPD is present, or on later stages, when planets dominate the system, and the disk has already dispersed. The intermediate phase, in which both planets and a substantial gaseous disk coexist, remains largely unexplored. This stage corresponds to the transition from gas-rich (Class II systems) to gas-poor debris disks\footnote{Also referred to as exo-Kuiper belts or Class III disks.}, with typical ages of $\sim$1–10 Myr for Class II objects and $\sim$10–100 Myr (or longer) for debris disk systems. For a comprehensive review of PPD evolution, see \citet{Williams&Cieza2011} and \citet{Armitage2020}. 

During the gas-rich phase, PPDs commonly exhibit dusty, ring-like substructures associated with gas-pressure maxima, often linked to the presence of planets \citep{Andrews2020}. These features are likely precursors of debris disks, and any modification of their architecture would influence the subsequent evolution once the gas has dispersed. Environmental perturbations, such as stellar flybys or infall events, can alter these structures in ways that persist into later evolutionary stages, including misalignments between debris disks and their host stars \citep{Hurt&MacGregor2023}. In particular, a specific class of debris disks --- narrow and eccentric rings --- might originate from such dynamical perturbations during the PPD phase. Narrow and eccentric debris disks are not rare or exotic, comprising approximately $5$–$10$\% of the resolved systems observed to date \citep{Hughes+2018, Esposito+2020, ARKSII, ARKSVI}. This relatively high frequency may indicate a shared underlying mechanism rather than stochastic or chaotic planetary evolution. These systems are difficult to reproduce within standard secular evolution models, since perturbing an axisymmetric debris disk with an eccentric planet will result in a broad disk with non-coherent eccentricities among the bodies that compose it \citep{Faramaz+2014,Pearce+2014}. This suggests that their eccentricities may have been imprinted while the gas was still present \citep{Kennedy2020}. Notable examples include the debris disks around HR~4796A \citep{Rodigas+2015}, Fomalhaut \citep{MacGregor+2017}, HD~202628 \citep{Faramaz+2019}, and HD~53143 \citep{MacGregor+2022}.

In this paper, we explore the consequences of environmental interactions between PPDs and both stellar flybys and infall events. We aim to characterize the dynamical imprints that these perturbations leave on the disk and embedded planets, with particular emphasis on the dust component. We further assume that the dust distribution traces the locations where planetesimals are likely to form and persist through the debris disk phase, thereby linking to observational signatures at later stages. The paper is structured as follows: Section~\ref{sec:methods} describes the numerical methods and simulation setup. Section~\ref{sec:results} presents our main results. Section~\ref{sec:discussion} discusses their implications. We conclude in Section~\ref{sec:conclusion}.

\section{Methods}
\label{sec:methods}

We considered a representative Class II PPD architecture, as exemplified by the PDS~70 system. This choice is based on the simplest architectural scenario and on the absence of any apparent signal of environmental interactions. We emphasize that our study does not adopt the specific parameters of PDS~70, but rather focuses on its general structural characteristics. In particular, we modeled a system consisting of a single central star, a disk \citep[as reported by][]{Keppler+2019}, and a pair of planets in MMR \citep[as suggested by][]{Keppler+2018, Haffert+2019, Bae+2019, Wang+2021}. 

We performed three-dimensional hydrodynamical simulations using the {\texttt Phantom} smoothed particle hydrodynamics (SPH) code \citep{PricePH+2018b} to model the interaction between our Class II system and both infalling cloudlet gas and stellar flybys. The simulations considered gas and dust SPH particles and employed the dust-as-particles (or two fluid) approach to model gas–dust coupling \citep{Laibe&Price2012a, Laibe&Price2012b, Mentiplay+2020,Price&Laibe+2020}, thereby self-consistently including the back-reaction of dust on the gas. 

\subsection{Fiducial disk setup}\label{sec:disk_setup}

We represented the stars and planets as sink particles \citep{Bate+1995}, adopting a central star mass of $2\ M_\odot$ and an accretion radius of 1 au. We included two planets in an MMR configuration, initialized in a coplanar 2:1 MMR. The planet's orbital parameters to achieve the MMR configuration are listed in Table \ref{tab:planet_params}. 

\begin{table}
\centering
\begin{threeparttable}
\caption{Disk parameters utilized in constructing the simulation.}
\begin{tabular}{lccc}

\hline
\hline

Name & Unit & Inner planet & Outer planet \\
\hline
Mass & $M_{\rm jup}$ & 1.00 & 1.00 \\
$a$ & au & 53.74 & 85.40 \\
$e$ & - & 0.24 & 0.06 \\
$i$ & degrees & 0.00 & 0.00 \\
$\omega$ & degrees & 194.66 & 191.98 \\
$\Omega$ & degrees & 0.00 & 0.00 \\
$\nu$ & degrees & 136.02 & 90.63 \\
$r_{\rm acc}$ & au & 2.70 & 4.80 \\

\hline
\hline

\end{tabular}

\begin{tablenotes}
\small
    \item \textbf{Notes.} The planetary orbital elements: semi-major axis ($a$), eccentricity ($e$), inclination ($i$), argument of periapsis ($\omega$), longitude of the ascending node ($\Omega$), and true anomaly ($\nu$). $r_{\rm acc}$ represents the accretion radius, which is approximately 1 Hill radius for a circular orbit.  
\end{tablenotes}

\label{tab:planet_params}
\end{threeparttable}
\end{table}

The disk is modeled using gas and dust SPH particles. The gas disk is initialized with $10^6$ particles in Keplerian rotation, a total mass of $10^{-3}\ M_\odot$, and radial \rev{extent} from $R_{\rm in}=120$~au to $R_{\rm out}=220$~au. The surface density follows a power-law profile, $\Sigma \propto r^{-1}$, while the disk is vertically isothermal with a temperature profile $T \propto r^{-0.5}$. These conditions yield a vertical aspect ratio of 0.105 at $R_{\rm in}$ and 0.122 at $R_{\rm out}$. We adopt an SPH shock viscosity parameter of $\alpha_{\rm AV}=0.28$, which corresponds to a mean Shakura–Sunyaev viscosity \citep{Shakura&Sunyaev73} of $\alpha_{\rm SS}\approx 3\times10^{-3}$.

The dust disk is initialized with the same spatial distribution and velocity field as the gas. An initial global dust-to-gas mass ratio of 0.01 is used, with a total of $10^5$ SPH particles split into five grain-size species, each represented by $2\times10^4$ particles. The grain sizes are spaced logarithmically, $s_i = \{9.48, 34.1, 122, 440, 1580\}\ \mu{\rm m}$. 

To quantify the degree of dust–gas coupling, we use the dimensionless Stokes number \citep{Birnstiel+2016}, ${\rm St}$, defined as
\begin{equation}
    {\rm St} = \sqrt{\frac{\pi}{8}\frac{GM}{r^3}}\frac{\rho_s s}{\rho_g c_{\rm s}},
\end{equation}
where $G$ is the gravitational constant, $M$ is the mass of the central star, and $r$ is the radial distance. The quantity $\rho_s$ denotes the intrinsic dust grain density, assumed to be $3\ {\rm g\ cm^{-3}}$ for all species; $s$ is the grain size, $\rho_g$ is the gas density, and $c_{\rm s}$ is the isothermal sound speed. The corresponding Stokes numbers for the different grain species are ${\rm St} \approx \{0.03, 0.13, 0.50, 1.70, 6.10\}$ at $R_{\rm in}$ and in the mid-plane.

The simulations are evolved for 100 outer-planet orbital periods ($T_{\rm out}$), or equivalently, 58870 years. This duration ensures that the dust species relax and form a well-defined dusty ring, thereby establishing a stable central cavity that hosts the two planets. The Stokes \rev{numbers evolve} for each grain species, and they are ${\rm St}_{0} \approx \{0.11, 0.35, 1.25, 5.60, 16.55\}$ at $R_{\rm in}$ and in the mid-plane by the end of this period. After 100 $T_{\rm out}$, a single external perturber --- either an infalling cloudlet or a stellar flyby --- is introduced. We also performed a control simulation that included the disk and planets without any perturbations.

\subsection{Environmental perturbers}

For the infall simulations, we followed the infall prescription of \citet{Calcino+2025}, modeling the infalling gaseous cloudlet as an ellipsoid of revolution with semi-major and semi-minor axes $A$ and $B$, respectively. We adopt $A = 1000$~au and $B = 250$~au, with a total mass of $10^{-3}\ M_\odot$ distributed among $10^6$ additional SPH gas particles. The ellipsoid is placed on a parabolic orbit, with its centre initially at 2250~au, resulting in a minimum approach distance of 440~au ($2R_{\rm out}$) on a coplanar, prograde trajectory relative to the disk plane and rotation. The velocity of each particle is initialized to the local free-fall velocity along the parabolic trajectory, with zero internal velocity dispersion within the ellipsoid.    

For the flyby simulations, we follow the stellar flyby prescription of \citet{Cuello+2019}. The \rev{perturber} star is modeled as a sink particle with a mass of $1\ M_\odot$ and an accretion radius of 1~au. It is initially placed on a parabolic orbit at a distance of 4400~au, which is ten times the closest approach distance of 440~au. The velocity of the sink particle is also initialized to the local free-fall velocity along the parabolic trajectory.

\begin{figure*}
\centering
\begin{center}
    \includegraphics[width=.9\textwidth]{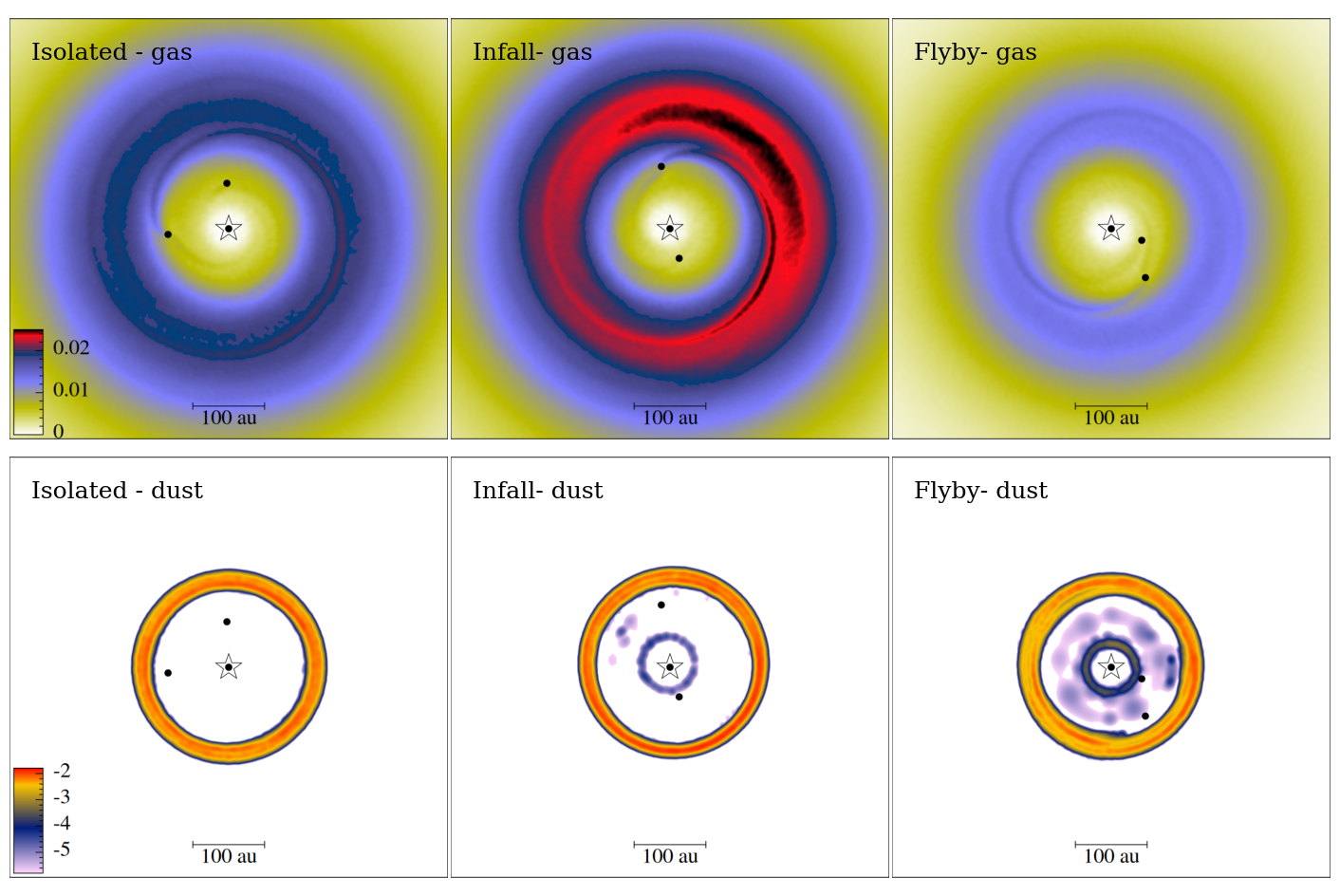} 
    \caption{Gas and dust surface density distributions at $300\ T_{\rm out}$. The top and bottom rows show the gas and combined dust (all grain species) surface densities, respectively. The three columns correspond to the isolated (left), infall (middle), and flyby (right) cases. The gas surface density is displayed on a linear scale, while the dust surface density is shown on a logarithmic scale, both in units of g cm$^{-2}$. The black dots indicate the positions of the planets, with their sizes corresponding to the adopted beam sizes.}
    \label{fig:Gas_panels}
\end{center}
\end{figure*}

\section{Results}
\label{sec:results}

After introducing either the infalling cloudlet or the stellar flyby, the simulations were evolved for an additional 200 $T_{\rm out}$, giving a total evolutionary time of 300 $T_{\rm out}$. We refer to the control run without external perturbations as the ``Isolated'' case, while the perturbed configurations are denoted as ``Infall'' and ``Flyby'', respectively. The resulting changes in the system architecture are primarily reflected in the gas and dust morphology, as well as in the planetary accretion rates, which are discussed below. \rev{Additionally, Appendix~\ref{app:pl} presents the evolution of the planetary orbital elements. As these parameters remain either unchanged or very similar to those in the Isolated case, they are not discussed further in the main text and are included in the appendix for completeness.}

\subsection{Gas disk}\label{sec:gaseous_disk}

\begin{figure*}
    \centering
    \begin{minipage}[t]{\columnwidth}
    \includegraphics[width=.95\textwidth]{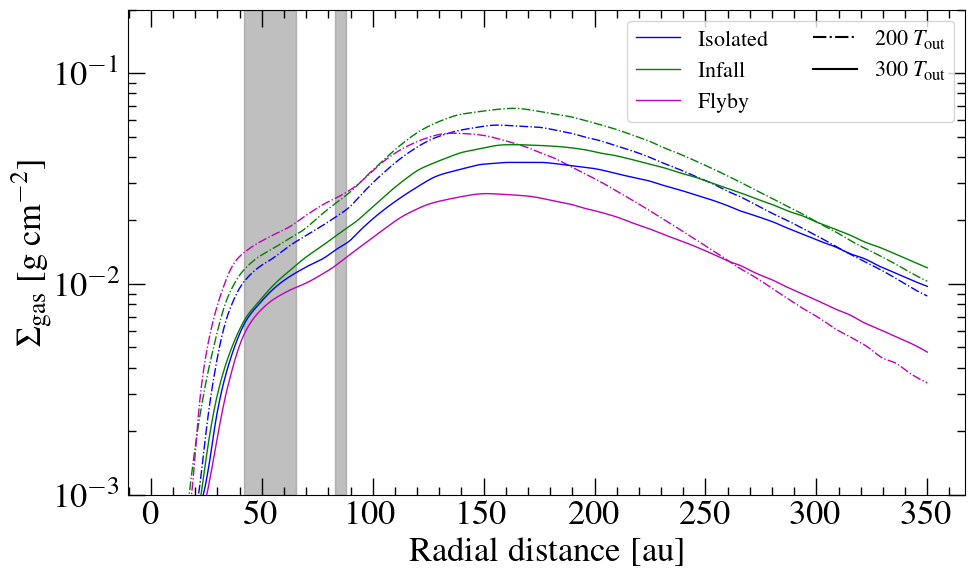} 
    \caption{Azimuthally averaged gas surface density profiles for the Isolated, Infall, and Flyby cases, shown in blue, green, and magenta, respectively. The segmented and solid lines represent evolving times of 200 and 300 $T_{\rm out}$, respectively. The shaded areas show the planets' radial excursions, determined by their pericenter and apocenter distances.}
    \label{fig:Gas_profile}
    \end{minipage}\hfill
    \begin{minipage}[t]{\columnwidth}
    \includegraphics[width=.95\textwidth]{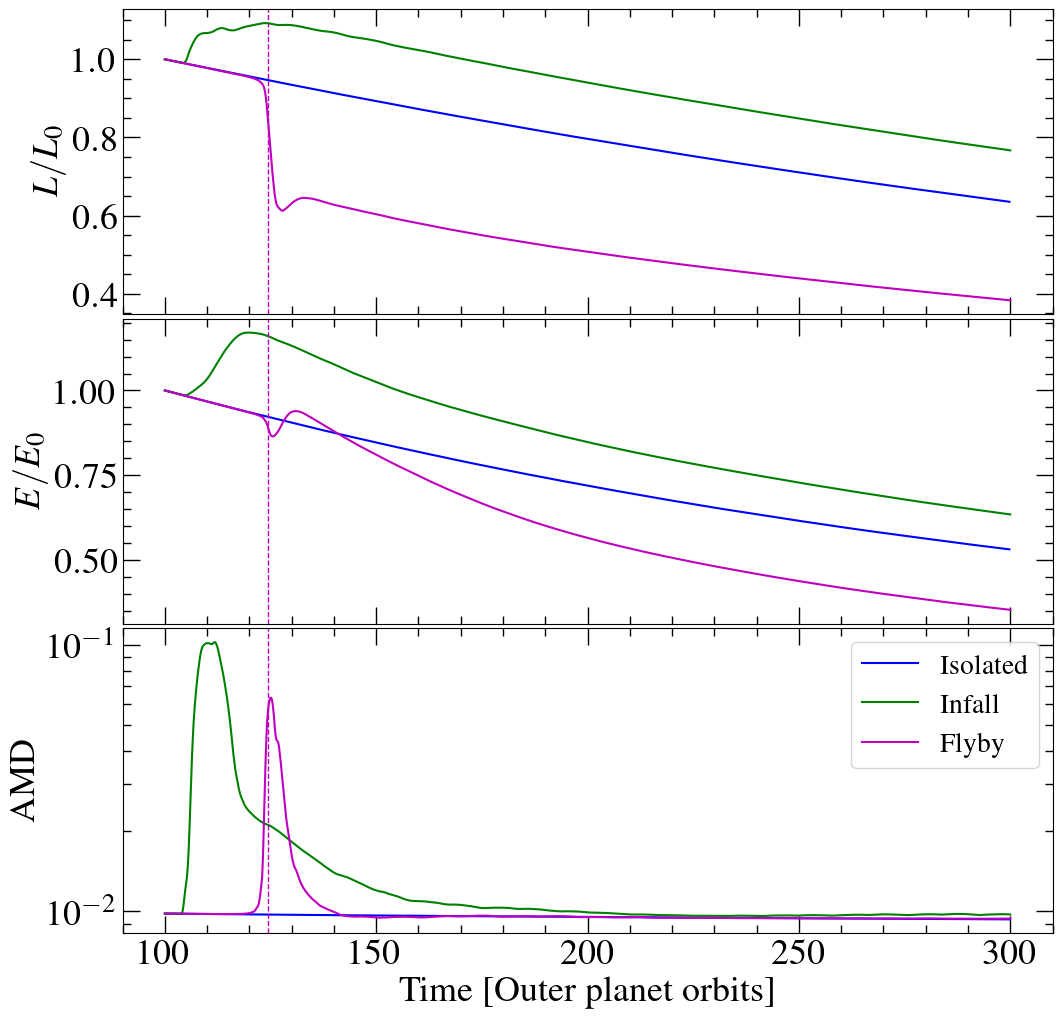} 
    \caption{Time evolution of the total angular momentum (top panel), total energy (middle panel), and angular momentum deficit (bottom panel) of the gas disk within 350 au. The color coding is the same as in Fig.~\ref{fig:Gas_profile}. The total angular momentum and total energy are normalized to their initial values. The vertical magenta-segmented line marks the time of closest approach of the stellar companion.}
    \label{fig:Gas_quantities}
    \end{minipage}
\end{figure*}

Figures \ref{fig:Gas_panels} and \ref{fig:Gas_profile} show the gas morphology and the corresponding azimuthally averaged surface density profiles for the three simulated cases after 300 $T_{\rm out}$. In all configurations, the gas component of the disk exhibits spiral arms, a characteristic outcome of planet–disk interactions \citep[e.g.][]{Ward1986,Rafikov2002}. 
\rev{However, their amplitudes differ significantly: the Infall case exhibits the largest surface-density enhancement, with a peak value of  $\sim 20\%$ and $\sim 70\%$ higher than those in the Isolated and Flyby cases, respectively, producing a pronounced azimuthal gas overdensity.} These trends are consistent with the global impact of each perturbation. The infalling cloudlet replenishes the disk by supplying additional mass, thereby increasing its angular momentum and energy, as discussed below. In contrast, the stellar flyby tidally truncates the disk, leading to the opposite behavior. This difference is quantified in Fig.~\ref{fig:Gas_profile}. 

In the Infall case, the disk is denser beyond the orbit of the outermost planet, whereas in the Flyby case, the outer regions are depleted. At $200\ T_{\rm out}$, both cases exhibit similar surface densities at 100~au; however, at 350~au, the infall case is denser by a factor of $\sim 3$. By $300\ T_{\rm out}$, the surface density profile becomes smoother, with surface densities in the Infall case exceeding the Flyby case by $\sim 50\%$ at 100 au and $\sim 150\%$ at 350~au. Conversely, the flyby initially promotes gas transport across the orbit of the outer planet, as reflected in the surface density profile at $200\ T_{\rm out}$. At 25~au, the surface density in the Flyby case is $\sim 25\%$ higher than in the Infall case. This trend reverses by $300\ T_{\rm out}$, when the Infall case becomes $\sim 50\%$ denser at the same radius. 

To characterize the disk, we computed the total angular momentum, total energy, and angular momentum deficit (AMD) of all SPH gas particles within 350 au, since the disk viscously spreads beyond the initial $R_{\rm in}$ location at 120 au. We considered the central star as the origin to derive the mentioned quantities. The total angular momentum is defined as the sum over the individual contributions of each SPH gas particle, that is,
\begin{equation}
    L=\left|\sum_i  \vb{r}_i\times m_i\vb{v}_i\right| ,
\end{equation}
where $m_i$, $\vb{r}_i$, and $\vb{v}_i$ are the mass, position, and velocity of an SPH gas particle, respectively. While for computing the total energy,
\begin{equation}
    E=\sum_i m_i \left( \frac{\left| \vb{v}_i\right|^2}{2} - \frac{GM}{\left| \vb{r}_i\right|} \right).
\end{equation}

\rev{The AMD is a measure of the degree of dynamical excitation in a system, originally introduced by \citet{Laskar1997} to quantify deviations from a coplanar, circular configuration in planetary systems. The AMD for the gas disk is shown in the bottom panel of Fig.~\ref{fig:Gas_quantities}. It is computed following the normalization adopted by \citet{Chambers2001}, defined as
\begin{equation}
    {\rm AMD}=1-\frac{L}{\sum_i L_{{\rm circ},i}},
\end{equation}
being $L_{{\rm circ},i}$ the angular momentum of an SPH gas particle in a circular orbit at the same semi-major axis, computed as
\begin{equation}
   L_{{\rm circ},i}= GM\frac{m_i^{3/2}}{\sqrt{-2E_i}}.
\end{equation}}

The time evolution of the total angular momentum and total energy of the gaseous disk are shown in the top and middle panels of Fig.~\ref{fig:Gas_quantities}. Before discussing the specific effects of the Infall and Flyby scenarios, we first note the systematic decrease in angular momentum and energy observed in all cases. This behavior results from accretion onto the sink particles (the central star and planets), which progressively reduces the number of SPH particles contributing to the disk reservoir and, consequently, the computed global quantities.

The evolution of the total angular momentum shows opposite trends in the Infall and Flyby cases. In the Infall scenario, the angular momentum increases by nearly 10\% over $\sim$20 $T_{\rm out}$, whereas in the Flyby case it undergoes an abrupt decrease of about 30\%. These behaviors are consistent with the respective addition and removal of material from the disk, as discussed above. The evolution of the total energy follows a similar trend in the Infall case, increasing by nearly 20\%. In contrast, the Flyby case exhibits more complex behavior: an initial slight decrease is followed by a modest increase, then a further decrease towards the end of the simulation. This pattern can be interpreted as the combined effect of mass loss due to gravitational capture and the transient gravitational perturbation from the passing star, which excites the disk. Comparatively, the ratio between the removal and injection for the total energy is smaller than for the total angular momentum for the Flyby case; the loss of angular momentum is significantly larger than the transient and subsequent increase induced by the passing companion. This indicates that the flyby removes angular momentum from the disk more efficiently than it alters its total energy.

Both the Infall and Flyby cases show an increase in the AMD, but they differ in both amplitude and duration. In the Infall scenario, the AMD reaches nearly 10\% of the value corresponding to a circular and coplanar configuration, whereas in the Flyby case, it rises to just above 5\%. Thus, the peak AMD in the Infall case is roughly a factor of two larger, indicating a more dynamically excited disk for the studied scenarios. The duration of the excitation also differs significantly between the two scenarios. Although both cases exhibit an initial sharp increase followed by a decline after about 10 $T_{\rm out}$, the AMD in the Infall case requires an additional 70 $T_{\rm out}$ periods to return to values comparable to those of the Flyby simulation, slightly higher than those of the Isolated simulation. By contrast, the Flyby case relaxes to the same steady state in fewer than 5 $T_{\rm out}$. This longer relaxation timescale in the Infall scenario arises because the gas cloudlet continues to interact with the disk even after the first encounter, which corresponds to the AMD peak. As material continues to accrete onto the disk, the system remains dynamically perturbed for an extended period, and it can even reverse the decreasing trend in AMD, as shown in Fig. \ref{fig:Gas_quantities}, which begins to increase slightly at 250 $T_{\rm out}$.  This highlights a key difference between the two studied excitation mechanisms: a stellar flyby produces a short-lived perturbation ($\sim 10\ T_{\rm out}$), whereas infalling material triggers an initial disturbance that continues to modify the disk's AMD over a longer timescale ($\sim 70\ T_{\rm out}$), although its intensity decreases over time. Nevertheless, the AMD slowly increases by the end of the considered evolutionary time.

\subsection{Dust disk morphology}\label{sec:dust_morphology}

\begin{figure*}
\centering
\begin{center}
    \includegraphics[width=.9\textwidth]{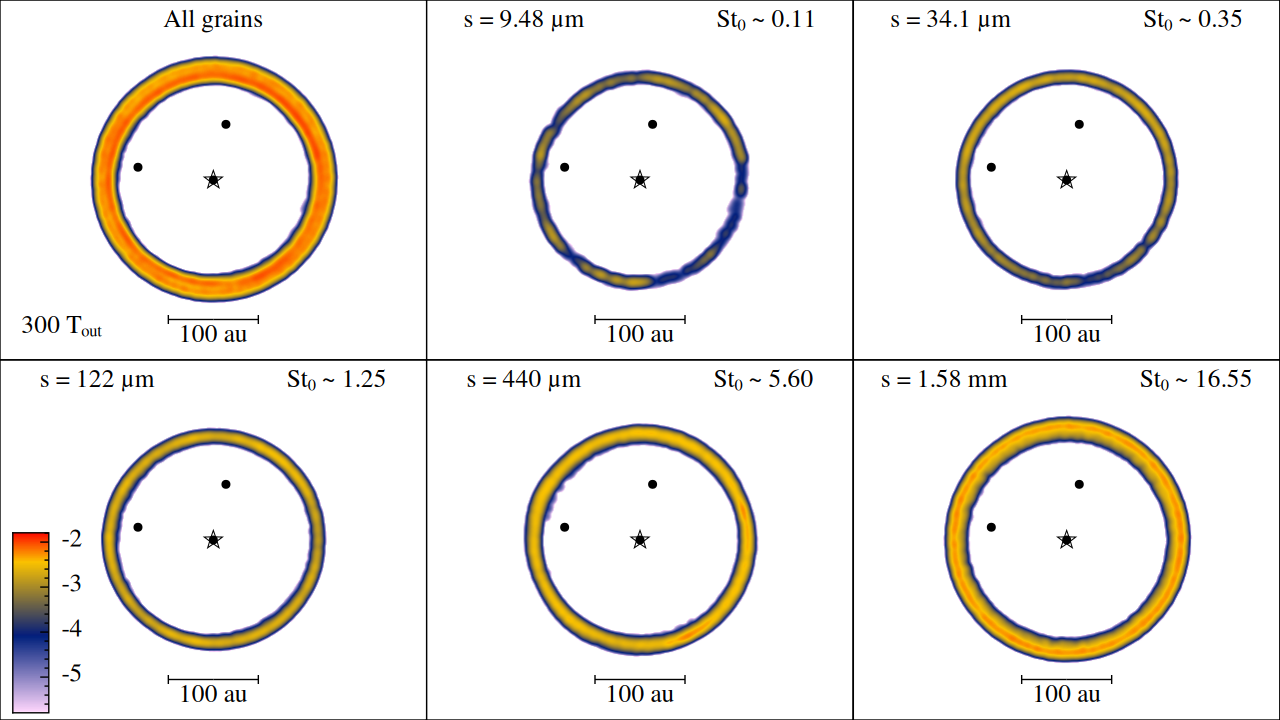} 
    \caption{Dust surface density after 300 $T_{\rm out}$ for the Isolated case. The first panel shows the combined contribution from all grain sizes (``all grains''), as also shown in Fig.~\ref{fig:Gas_panels}, while the remaining panels display the surface density of each of the five dust species separately. The surface density is presented on a logarithmic scale in units of g cm$^{-2}$. Black dots mark the planetary positions, with their sizes indicating the adopted beam size.}
    \label{fig:control_300}
\end{center}
\end{figure*}

\begin{figure*}
\centering
\begin{center}
    \includegraphics[width=.9\textwidth]{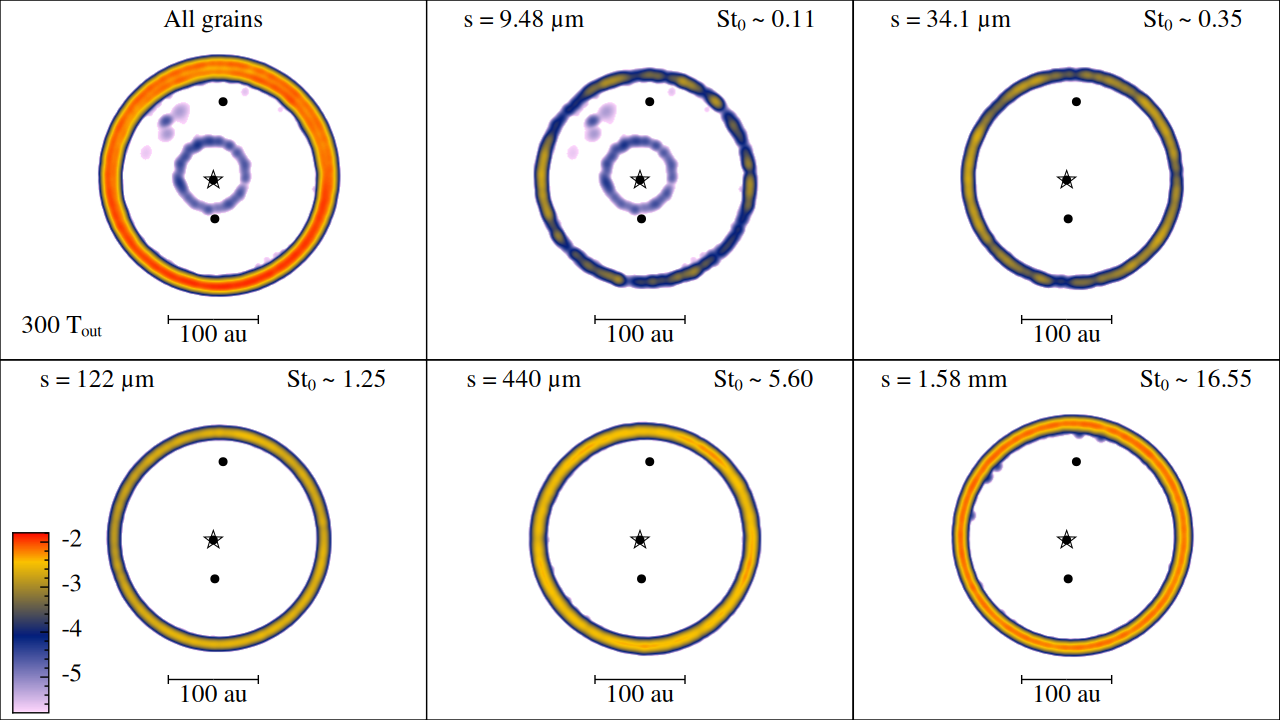} 
    \caption{Same as Figure \ref{fig:control_300} but for the Infall case}
    \label{fig:infall_300}
\end{center}
\end{figure*}

\begin{figure*}
\centering
\begin{center}
    \includegraphics[width=.9\textwidth]{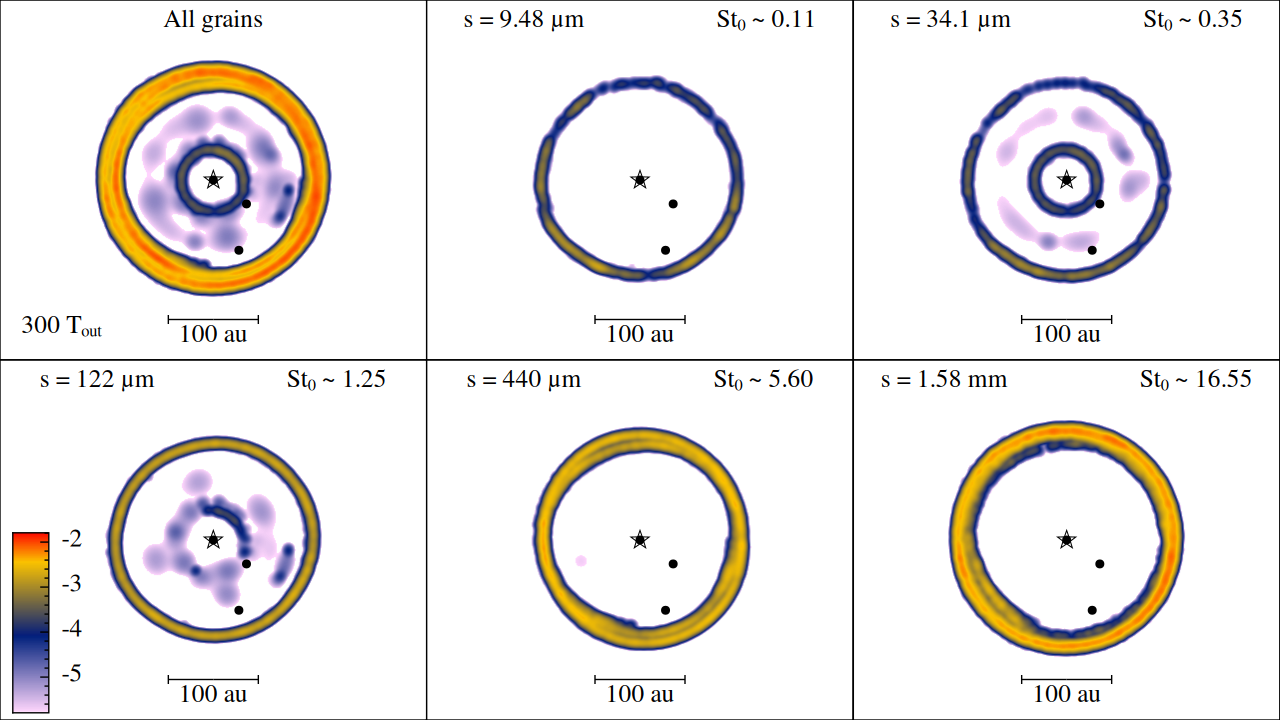} 
    \caption{Same as Figures \ref{fig:control_300} and\ref{fig:infall_300} but for the Flyby case}
    \label{fig:flyby_300}
\end{center}
\end{figure*}

Changes in gas dynamics affect dust evolution, depending on the degree of coupling between dust grains and the gas fluid. The Stokes number quantifies such a coupling, meaning dust grains with $\rm St \ll1$ and $\rm St \gg1$ are tightly coupled and decoupled with gas, respectively \citep[see][]{Birnstiel+2016}. Particularly, dust with ${\rm St}=1$ exchanges the most angular momentum with the gas. Our setup, described in Section \ref{sec:disk_setup} considered a range of grain sizes across the coupled and decoupled regimes at the initial state, i.e.  ${\rm St}_{0} \approx \{0.11, 0.35, 1.25, 5.60, 16.55\}$. Figures \ref{fig:control_300}, \ref{fig:infall_300}, and \ref{fig:flyby_300} show the resulting morphology of the dust disk for the Isolated, Infall, and Flyby cases, respectively, after 300 $T_{\rm out}$. 

The dust morphology in the Isolated case, shown in Fig.~\ref{fig:control_300}, exhibits the classical features expected for a Class II system in the simplest configuration: a central dust cavity surrounded by a single circular and smooth dust ring, with no prominent features radially or azimuthally for all dust grains. A cavity surrounded by a smooth dust disk is observed in continuum observations of systems like PDS~70, where planets have been confirmed within the dust cavity \citep{Keppler+2019}. The dust ring forms at the location of the pressure maximum generated by the outermost planet as it carves a gap in the disk, in agreement with theoretical predictions and numerical models \citep{Fouchet+2010, Pinilla+2012}. 

\rev{In the Isolated case, none of the dust species crosses the orbit of the outermost planet, as they become trapped at the pressure bump. Even the smallest grains, despite being strongly coupled to the gas, remain sufficiently decoupled to prevent significant inward drift across the planetary barrier. In the Infall and Flyby cases, however, the dust ring exhibits significant variations across grain sizes. In the Infall scenario, the dust ring remains relatively narrow across all grain species, but an additional secondary ring appears at smaller radii, interior to the inner planet's orbit, for the smallest grains. 
The additional gas supply, together with the initial impact of infalling material in the Infall case, overcomes the planetary barrier, allowing the smallest dust grains to reach the innermost regions of the disk. While in the Flyby case, the density waves induced by the passing star drive both gas and dust inward, allowing the dust population to enter the planetary orbits and reach the innermost regions.
}

\rev{Nevertheless, we caution against over-interpreting the location of the inner ring, which appears at $\sim 40$ au in our simulations. Although an inner ring is expected to form closer to the star, near the sublimation radius \citep{Eisner+2005,Andrews+2009}, its position in our models may be influenced by the adopted numerical setup, including the stellar sink radius and the orbit of the innermost planet. While the effect of the sink radius is likely negligible given its value of 1 au, the planetary configuration may play a more significant role in determining the ring's radial position, such as through resonance capture.} An inner ring is therefore expected to form, but probably at a different radial distance. Moreover, at least two distinct levels of eccentricity can be identified, as illustrated in the all grains panel of Fig.~\ref{fig:infall_300}. The combined dust distribution suggests an azimuthal asymmetry: in one sector, all dust remains concentrated in a narrow band, whereas on the opposite side it spreads out, forming a double-ring structure. This pattern arises from the impact of the infalling gas cloudlet in a specific azimuthal region of the disk, which coincides with the sector where all the dust remains more confined.

In the Flyby case, gas is pushed inward as it crosses the outermost planetary orbit, as discussed in the previous section. This behavior affects the dust distribution by allowing grains to cross planetary orbits and migrate into the inner regions of the disk. Unlike the Infall case—where only the smallest grains produce an additional interior dust ring to the inner planet—the Flyby scenario forms such an inner ring for grains with ${s}=\{34.1,122\}\ \micro$m, while it is absent for the smallest grain size. This behavior suggests that radial dust migration is not driven solely by gas hydrodynamical effects; rather, the gravitational perturbation from the passing star compresses the dust distribution radially, forcing grains to migrate inward. This is supported by observations, but in the context of stellar multiplicity \citep{Manara+2019}. Furthermore, the all-grains panel shows a broader outer ring than the other two simulated cases. In addition, a double-ringed structure can be identified in the dust distribution for grains with $s=440\ \micro$m, which likely results from the radial dynamical excitation triggered by the encounter. 

\subsection{Dynamical excitation of dust grains}\label{sec:dust_excitation}

In Sect.~\ref{sec:gaseous_disk}, we quantified the level of dynamical excitation in the gaseous disk and, by extension, in the overall disk structure, given that gas is approximately two orders of magnitude more abundant than dust. However, dust does not necessarily respond to perturbations in the same way as the gas, as seen above. It is therefore necessary to independently quantify the level of dynamical excitation in the dust component. Unlike the analysis in Sect.~\ref{sec:gaseous_disk}, where the disk was considered out to 350 au, we characterized the dust distribution within the radial range 80--150 au, corresponding to the location of the outer dust ring. The eccentricity of the dust disk for each grain size is computed as the median of the magnitude of the eccentricity vector, namely
\begin{equation}
    e_{\rm disk} = {\rm median}\left(\left| \frac{ \vb{v}_i\times (\vb{r}_i\times \vb{v}_i)}{GM} - \frac{\vb{r}_i}{\left| \vb{r}_i\right|} \right|\right). 
\end{equation}
We additionally computed $\varpi_{\rm disk}$ as the median of the longitude of the pericentre of each particle. For completeness and direct comparison, we also compute the eccentricity of the gaseous disk within the same radial region. It is worth noting that $e_{\rm disk}$ is derived from the intrinsic dynamical properties of particles, which can be easily obtained through numerical simulations. However, an alternative value, $e_{\rm geo, disk}$, can be derived by fitting an ellipse to the dust morphology; this aligns with the eccentricity derived from observations of dusty rings. In the case of narrow rings, $e_{\rm disk}$ and $e_{\rm geo, disk}$ only converge if the longitude of pericentre or each particle, $\varpi_i$, exhibits a low dispersion. This is important since, from a geometric fitting, it is possible to infer the physical eccentricity distribution. The dispersion was quantified by measuring the median absolute deviation $\sigma_{e}$ and $\sigma_{\varpi}$ for the eccentricity and the longitude of pericentre, respectively.

\subsubsection{Eccentricity evolution}

The top row of Fig.~\ref{fig:disk_eccs} shows the resulting evolution of eccentricities for all dust species and SPH gas particles. The Isolated case remains nearly circular for both the gas and dust components with low dispersions. It is worth noting that the two planets are initially in a 2:1 MMR, with eccentricities of $\sim0.06$ and $\sim0.24$ for the outer and inner planets, respectively. Despite these relatively large planetary eccentricities, the dust disk remains essentially circular, while the gas disk shows only a marginal response, with its eccentricity increasing to $\sim0.02$.

In the Infall case, a clear distinction emerges between coupled and decoupled dust grains. The coupled grains (${\rm St}\ll1$) exhibit similar evolutionary trends: their eccentricities peak at $\sim0.25$ and subsequently decline gradually to $\sim0.05$ by the end of the simulation. After the peak, their eccentricities remain systematically higher than those of the gas by approximately 20\%. The two simulated decoupled grain populations (${\rm St}\gg1$) display different behavior. In general, their response to variations in gas eccentricity is delayed, with the delay increasing as the Stokes number increases. Consequently, while the coupled grains begin to decrease their eccentricities after the peak, the decoupled grains continue to increase theirs. The grains with $s=440\ \micro$m reach a peak eccentricity of $\sim0.20$, whereas those with $s=1.58$~mm attain a lower peak of $\sim0.13$. Despite differences in eccentricity median values, the dispersion of all grains remains low and similar. These results confirm the scenario described in the previous section, in which multiple dust eccentricities coexist simultaneously for different grain sizes. This effect is particularly noticeable from $175\ T_{\rm out}$ until the end of the simulation, where the largest grains retain the highest eccentricity among dust species despite having reached the smallest peak values earlier in the evolution. By the end of the simulation, as in the AMD trend change, gas begins to acquire eccentricity, and consequently, the remaining grains do as well. The behavior of the largest grains suggests that dust grains with larger Stokes numbers will maintain the highest eccentricities despite the circularization of gas.

The large eccentricities exhibited in the Infall case contrast with the AMD analysis discussed in Sect.~\ref{sec:gaseous_disk}, where the gaseous disk appears approximately relaxed by the end of the simulation. The apparent discrepancy can be explained by the different spatial regions considered in the two analyses. The AMD calculation includes the entire disk, whereas the eccentricity measurement focuses on the inner region. As a result, a fraction of the inner disk now appears dynamically excited, but this does not significantly affect the global AMD. Because dust preferentially accumulates in this region, it effectively traces the residual dynamical excitation, distinguishing an eccentric inner disk region adjacent to the outer planet from a more dynamically relaxed outer disk. However, regions beyond 350 au still contain infalling residual gas from the cloudlet, which is expected to exhibit increased eccentricity \citep[e.g.][]{Kuffmeier+2021}. These regions are likely to trigger the increases observed in dust eccentricities and AMD by the end of the evolutionary time.

In the Flyby case, the dust and gas components exhibit a similar dynamical evolution. Their eccentricities initially increase to $\sim0.1$, before decreasing to values below 0.05, where they subsequently remain. Their dispersions are also slow, indicating a coherent eccentricity distribution. Notably, by the end of the simulation, the disk components still retain a non-negligible level of dynamical excitation, except for $s=440\ \micro$m, which experiences a modest increase. 

\subsubsection{Longitude of pericentre evolution}\label{sec:varpi_evol}

\begin{figure*}
\centering
\begin{center}
    \includegraphics[width=1\textwidth]{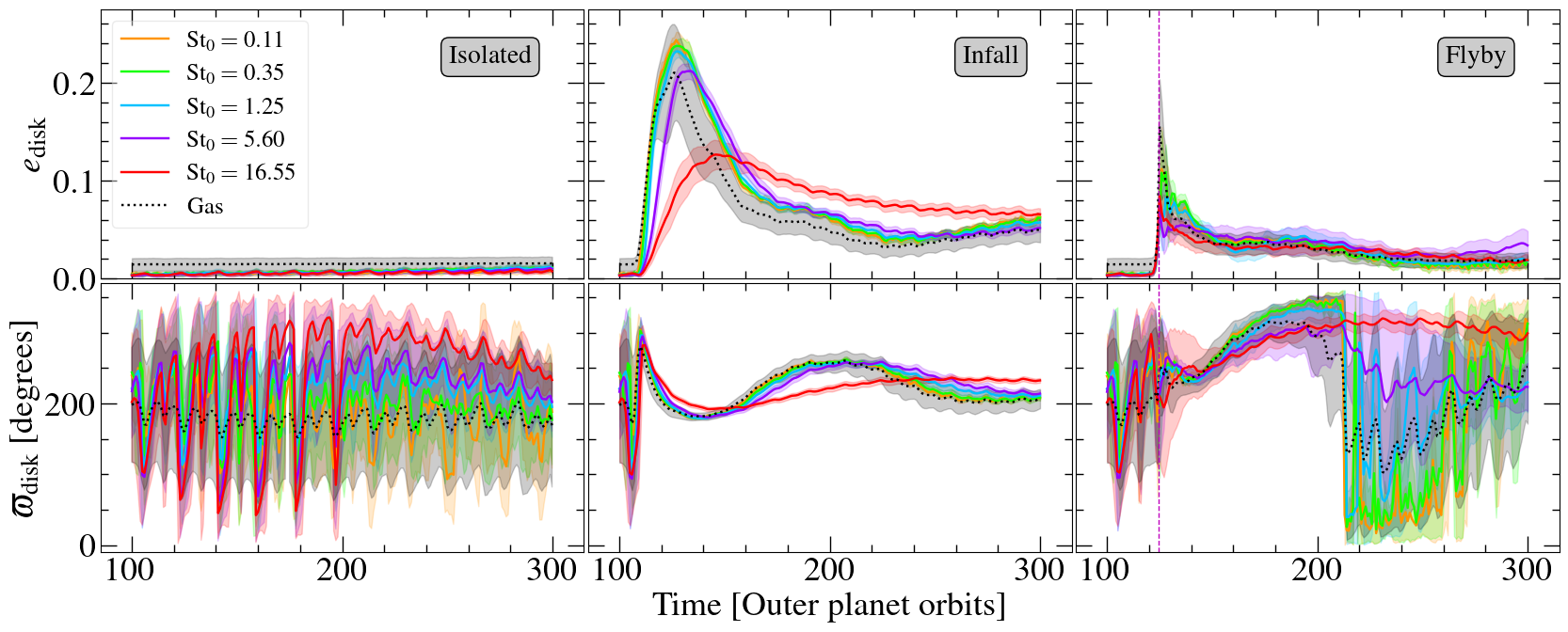} 
    \caption{Evolution of the median eccentricity and longitude of pericentre of SPH dust and gas particles within 80–150 au. The left, middle, and right panels correspond to the isolated, infall, and flyby cases, respectively. The top and bottom rows show the eccentricity and longitude of pericentre, respectively. The eccentricities of the different dust species are color-coded, while the gas disk is represented by a black dotted line. In the right column, the vertical black line marks the time of closest approach of the stellar companion.}
    \label{fig:disk_eccs}
\end{center}
\end{figure*}

The bottom row of Fig.~\ref{fig:disk_eccs} shows the evolution of the longitude of pericentre for all dust species and SPH gas particles. In the Isolated case, $\varpi_{\rm disk}$ exhibits a broad distribution spanning nearly the full range of $0$–$360\degree$, indicative of a lack of preferential orientation and an incoherent distribution. The introduction of perturbers leads to a realignment of these values, again highlighting the distinction between coupled and decoupled dust populations.

In the Infall case, $\varpi_{\rm disk}$ exhibits a high degree of coherence across both dust and gas species as eccentricity increases. The values are confined to the range $200$–$220\degree$, with negligible dispersion. The $\varpi_{\rm disk}$ of the strongly coupled dust grains closely follows that of the gas. In contrast, the $s = 440\ \mu$m and $s = 1.58$~mm grains show a delayed response, similar to their eccentricity evolution, with the largest grains exhibiting the longest lag. This behavior, together with the outlying eccentricity of the largest grains, supports the coexistence of multiple eccentric and aligned populations within the same dust ring (Fig. \ref{fig:infall_300}) for at least $\sim 125\ T_{\rm out}$. By the end of the simulations, both $e_{\rm disk}$ and $\varpi_{\rm disk}$ show a tendency toward convergence, potentially leading to the formation of a single, narrow, eccentric dust ring.

In the Flyby case,  the $\varpi_{\rm disk}$ evolution can be divided into two phases. Immediately after the stellar encounter, $\varpi_{\rm disk}$ becomes relatively coherent across all species. This phase lasts for $\sim 60\ T_{\rm out}$. Subsequently, at $\sim 210\ T_{\rm out}$, the gas $\varpi_{\rm disk}$ shifts from $\sim 300\degree$ to $\sim 100\degree$. The strongly coupled grains, as well as the $s = 440,\mu$m species, follow this transition, accompanied by a marked increase in dispersion comparable to that of the Isolated case. In contrast, the $s = 1.58$ mm grains remain relatively coherent for the rest of the evolutionary time. The resulting spread in $\varpi_{\rm disk}$ leads to an overall broadening of the dust disk, as illustrated in Fig.~\ref{fig:flyby_300}.

To summarize the final states of $e_{\rm disk}$ and $\varpi_{\rm disk}$, Fig.~\ref{fig:e_free} shows the distribution of the largest dust grains in the $hk$-plane for the three simulated cases, where $(h,k) \equiv (e\cos\varpi,\ e\sin\varpi)$. The Infall case shows a single, coherent, and eccentric distribution, whereas the Flyby case is characterized by a less eccentric, moderately coherent population, together with an additional, more dispersed component. We discuss the implications for subsequent evolutionary stages, namely the debris disk phase, in Sect.~\ref{sec:debris_disk}.

\subsection{Planetary accretion}

\begin{figure*}
    \centering
    \begin{minipage}[t]{\columnwidth}
    \includegraphics[width=.95\textwidth]{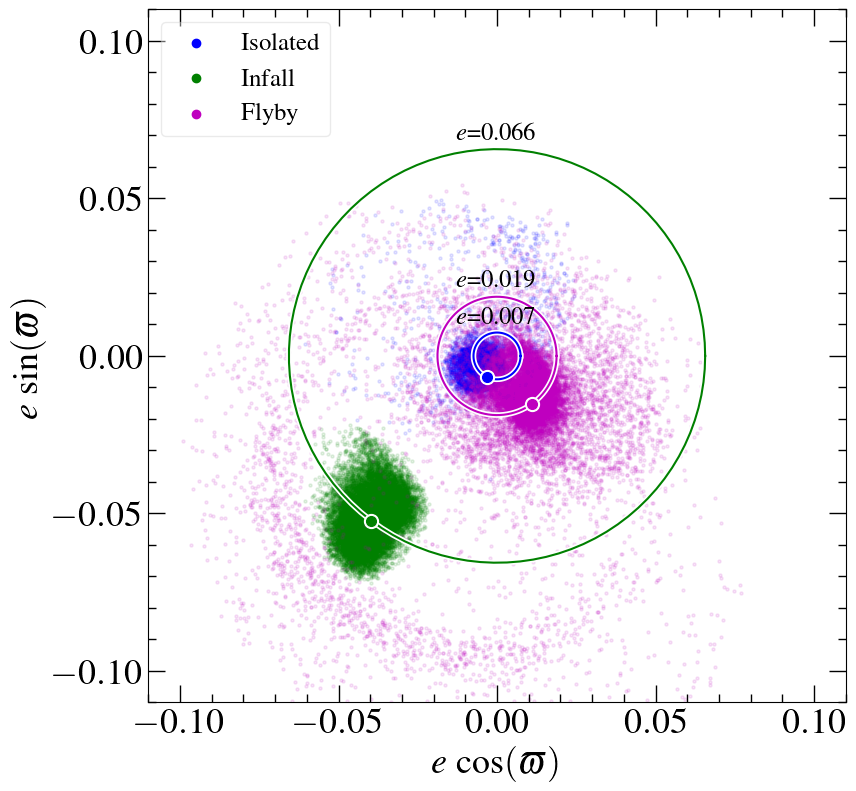} 
    \caption{Final distribution in the $hk$-plane for the dust grains with the highest Stokes number (i.e. $s=1.58$ mm) in all simulated cases. The color coding is the same as in Figs.~\ref{fig:Gas_profile}, and \ref{fig:Gas_quantities}. The colored circles indicate the median eccentricity of each distribution, while the centroids are marked by thicker points. }
    \label{fig:e_free}
    \end{minipage}\hfill
    \begin{minipage}[t]{\columnwidth}
    \includegraphics[width=0.95\textwidth]{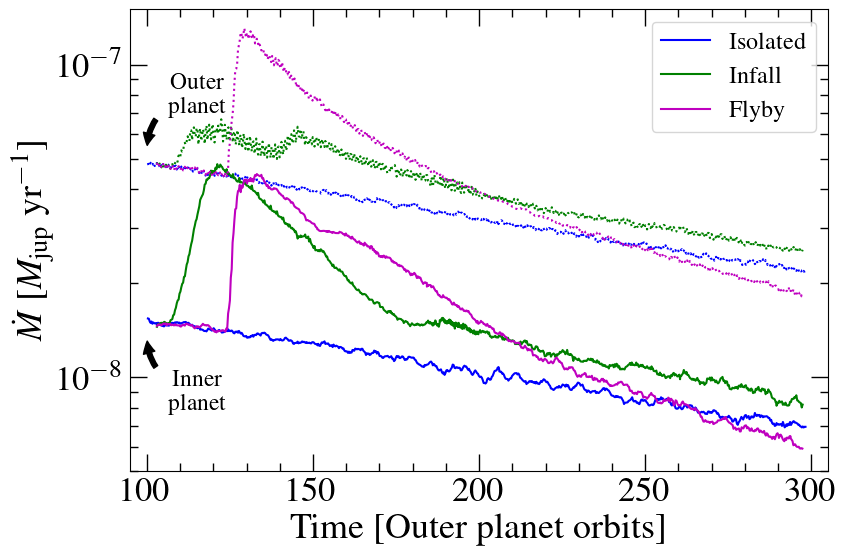} 
    \caption{Time evolution of the accretion rates onto the inner and outer planets, shown with solid and dotted lines, respectively. The color coding is the same as in Figs.~\ref{fig:Gas_profile}, \ref{fig:Gas_quantities}, and \ref{fig:e_free}.}
    \label{fig:Planet_mdot}
    \end{minipage}\hfill
\end{figure*}

Planetary accretion is also affected by infall and flyby events. Figure~\ref{fig:Planet_mdot} shows the time evolution of the accretion rates for the two planets in the three simulated cases. In the Isolated case, the outer planet exhibits a higher accretion rate than the inner planet, as expected, since it accretes directly from the surrounding gas disk. During both the flyby and infall events, the accretion rate of the inner planet increases significantly, reaching peak values comparable to the steady-state accretion rate of the outer planet and nearly doubling its initial value. This enhancement is followed by a gradual decline. The outer planet, however, responds differently in the two perturbed scenarios. In the Flyby case, its accretion rate increases by a factor of $\sim 2.5$, whereas in the Infall case, the increase is more modest, reaching only $\sim 1.25$ times its initial value.

\rev{The latter trend indicates a reduction in the potential accretion onto the outermost planet in the Infall case, whereas the accretion behavior of the innermost planet remains broadly similar to that in the Flyby case. This reduction may be attributed to the additional angular momentum injected into the disk by the infalling material. As a consequence, gas originating from both the disk and the cloudlet acquires higher angular velocities, making it more difficult for the outer planet to efficiently accrete the mixed material. This interpretation is consistent with the results of \citet{Kuffmeier+2021}, who found that retrograde infall enhances the accretion rate onto the central star. In a retrograde configuration, the relative angular velocities of the disk and cloudlet are reduced, leading to a net loss of angular momentum. By analogy, such a configuration would likely facilitate accretion onto the outermost planet in our scenario, in contrast to the suppression observed for prograde infall.}

Overall, in the Infall case, the accretion rates of both planets remain higher than in the isolated case and decrease at a similar rate after $200\ T_{\rm out}$. By contrast, in the Flyby case, both planets exhibit the highest accretion rates during the first $\sim 200\ T_{\rm out}$ compared to the other cases, after which they drop below the Infall case and fall beneath the Isolated case by $\sim 250\ T_{\rm out}$.

These results have important implications for planetary evolution when considered together with the processes discussed in the previous sections. In the Flyby case, the accretion rates increase for both planets; however, the accreted material is not limited to gas. As shown in Sect.~\ref{sec:dust_morphology}, the stellar encounter accelerates radial dust migration by truncating the disk and increasing the steepness of the gas surface density, and forcing grains to cross planetary orbits. This suggests that the planets may accrete a larger fraction of solid material than in the Isolated and Infall cases, where dust predominantly remains trapped in the outer ring. Consequently, the Flyby case may lead to enhanced solid enrichment of the planets, potentially affecting their chemical composition and metallicity. In the Infall case, the evolution of planetary accretion rates is more counterintuitive: the inner planet shows the largest increase. This implies that the inner planet would accrete a substantial fraction of the additional material delivered by the infalling gas cloudlet during the initial encounter, comparable to that accreted by the outer planet. \rev{Nevertheless, this could change if a retrograde infall event is considered.}

\begin{figure}
\centering
\begin{center}

\end{center}
\end{figure}

\section{Discussion}
\label{sec:discussion}

Considering environmental perturbers, such as infalling gaseous material or stellar flybys, during the late stages of planet formation can significantly affect the resulting planet–disk architecture. Although our study explored a limited set of parameters, the results reveal clear structural and dynamical signatures that may be extrapolated to a broader range of configurations. Moreover, the signatures of these interactions may persist into later evolutionary stages and remain observable in evolved systems during the debris disk phase.

\subsection{Impact of initial conditions}

A different choice of parameters for both the planet–disk system and the external perturbers can significantly modify the resulting configurations. In this work, we assumed coplanarity at all levels, namely between the planets, the disk, and the perturbers. In a non-coplanar configuration, the orbital planes of the planets and the disk may be altered. Such interactions could modify the planet–planet architecture and potentially disrupt mean-motion resonances \citep{Charalambous+2025}. In addition, the direction of the disk angular momentum may vary, as explored by \citet{Kuffmeier+2021} in the context of infall events.  The magnitude of this variation depends on both the disk rotational inertia and the strength of the perturbation. The disk inertia is primarily determined by its mass distribution and viscosity, while the perturber controls the amount of angular momentum injected or removed. In the case of infall, this depends on the mass, impact location, and direction of the accreting material, whereas in the flyby scenario, it is mainly controlled by the mass, passing star's pericentre distance, and orientation.

The prevalence of gas disk eccentricity was investigated by \citet{Ragusa+2018}, who identified two regimes in which the eccentricity of the disk may either be damped or amplified, depending on the planet-to-disk mass ratio, at the expense of the planet’s eccentricity. Their results indicate that eccentric modes can persist for at least $10^3$–$10^5$ planetary orbits; in our case, this corresponds to $\sim10^5$–$10^7$ yr. Additionally, this timescale is supported by \citet{Kuffmeier+2021}, who, based on hydrodynamical simulations, report that an eccentric gaseous disk persists for at least $1.5\times10^5$ yr. Thus, in an evolved system, the gas disk could retain a non-negligible eccentricity during its lifetime. However, these timescales are also sensitive to local disk conditions, such as pressure and viscosity \citep{Teyssandier&Ogilvie2016}. Infalling material can especially modify the disk pressure profile by increasing the total mass and altering its optical properties—effects that our model simplifies by assuming an isothermal equation of state. Furthermore, the aforementioned studies consider single-planet configurations, and, in the presence of an MMR, the outer planet's eccentricity is constrained, which can modify the evolution of the disk's eccentricity. Although gaseous eccentric features may dissipate on timescales of $\sim10^5$ years, dust particles with large Stokes numbers can preserve these signatures for longer periods and may remain observable in later evolutionary stages.

\subsection{Observational implications of environmental perturbations}

Our findings can be compared with observations of systems reported in the literature. In particular, multi-wavelength studies can be used to characterize dust grains with different Stokes numbers \citep[e.g.,][]{Alaguero+2025}, thereby testing the proposed solid segregation.

The HL~Tau system provides an illustrative example of an infall event. Classified as a Class I–II object \citep{Furlan+2008}, it exhibits a multi-gap and ringed structure in millimeter emission \citep{ALMA+2015}, consistent with the presence of embedded, nearly circular protoplanets \citep{Dipierro+2015}. At the same time, spiral arms have been associated with a past infall event observed in the molecular line HCO$^+$(3-2) \citep{Yen+2019}. The coexistence of these features suggests that the disk may have sufficient rotational inertia, that the angular momentum carried by the infalling material is relatively low, and/or that the dust grains traced by ALMA are coupled from the gas. In this case, the disk can maintain a largely axisymmetric structure despite environmental perturbations.

Among the observed systems with evidence of ongoing flyby interactions, ISO-Oph~2 and AS~205 are particularly notable for their substructure in continuum observations. (i) ISO-Oph~2, although interpreted as a wide binary system with a separation of $\sim 240$ au \citep{Cieza+2019}, could be better explained by a flyby scenario. Its morphology consists of two non-axisymmetric rings separated by a central cavity of $\sim 50$ au, together with a bridge connecting the two stellar components detected in $^{12}$CO emission \citep{Gonzalez-Ruilova+2020}. (ii) AS~205 is a triple system in which two components, AS~205N and AS~205S, are spatially resolved, with the southern component itself being a spectroscopic binary \citep{Eisner+2005}. A bridge connecting both components is also observed in scattered-light images \citep{Weber+2023}, with a projected separation of $\sim 168$ au. The southern component further exhibits a ring-like continuum emission with a semi-major axis of the order of 30 au \citep{Kurtovic+2018}.

The systems discussed above exhibit direct signatures of the environmental perturbations considered in this work. Detailed analysis of these systems should therefore focus on characterizing dust dynamics to assess the consistency between the observed solid distribution and our results. In particular, eccentric dust rings, together with multiple grain populations with distinct eccentricities, emerge as common characteristic signatures of both infall and flyby events. On the other hand, the main difference, despite the larger eccentricity reached by the Infall case, is the spread in the longitude of pericentre for most grains in the Flyby case. This may increase relative grain velocities and enhance mutual destructive collisions. 

Our findings can also be used as a predictive framework. The systems HD~100546 \citep{Perez+2020,Fedele+2021} and L1448~IRS3A \citep{Reynolds+2024}, for instance, show evidence of eccentric rings at millimeter wavelengths, which may point to past infall or flyby interactions. Such perturbations can imprint long-lasting structures on the disk that persist even after the gas has dissipated. This may provide a natural pathway to the formation of one of the most intriguing debris disk morphologies: narrow, eccentric rings, as observed in Fomalhaut-like systems.

\subsection{Fomalhaut-like debris disks}\label{sec:debris_disk}
Explaining narrow and eccentric debris disks within the framework of classical secular evolution is challenging. Perturbations of an initially axisymmetric disk by an eccentric planet are expected to produce a radially extended structure with non-coherent eccentricities among the constituent bodies. This, together with the significant fraction of observed narrow, eccentric debris disks, suggests an alternative formation mechanism that remains unknown. \citet{Kennedy2020} suggested these debris disks are born narrow and eccentric. More recently, \citet{Lovell+2025} supported this scenario using $N$-body simulations of the Fomalhaut system, showing that although an eccentric planet can sustain the current eccentric disk morphology, such a planet is unable to reproduce the disk's narrowness, suggesting that Fomalhaut's disk was born eccentric, i.e., it acquired its eccentricity from its progenitor PPD.
Accordingly to our findings, its origin may be linked to environmental perturbations, such as stellar flybys or infalling material occurring during the late stages of the PPD phase. Similar to the scenarios proposed by \citet{Lyra&Kuchner2013} and \citet{Lin&Chiang2019}, our work suggests that it may not be necessary to invoke eccentric planets to produce eccentric debris disks

To extrapolate our results to the debris disk phase, it is important to note that the dust origin differs between PPDs and debris disks. In debris disks, the dust population is not necessarily inherited directly from the PPD; instead, most small grains are produced through collisional cascades of larger bodies \citep[][]{Wyatt2008,Lohne+2008}. In this context, dust particles with larger Stokes numbers become particularly relevant, as they more readily inherit dynamical perturbations during the PPD phase, as described in Section \ref{sec:dust_excitation}, and they transfer their orbital properties to the fragments when they collide \citep{Wyatt+1999}.  This suggests that high-Stokes-number dust preferentially traces the regions where larger planetesimals are likely to form, which subsequently evolve into the debris disk.

As discussed in Section~\ref{sec:dust_excitation}, the Infall case produces the eccentric and coherent distributions needed to form narrow, eccentric debris disks. Nevertheless, this condition is necessary but not sufficient to sustain such a disk structure into the debris-disk phase. According to secular evolution theory \citep[see][]{Murray&Dermott1999}, such distributions precess around a fixed point in the $hk$-plane corresponding to the forced eccentricity, with a corresponding forced longitude of pericentre, both quantities being functions of the orbital properties of the present planets, such as their eccentricities, longitude of pericentres, and semi-major axis. Over long timescales, this precession can have two effects on particles depending on their properties and forced components values: 
\begin{enumerate}
    \item $(e_{\rm disk},\ \varpi_{\rm disk})$ differs from the forced components. This process randomizes the particles’ longitude of pericentre, resulting in a circular distribution of points around the forced eccentricity vector. Consequently, the dust disk attains a broader spatial extent, with its inner and outer edges linked to the minimum and maximum particle eccentricities, respectively.
    \item $(e_{\rm disk},\ \varpi_{\rm disk})$ aligns with the forced components.This process randomizes the particles’ longitude of pericentre; however, because their eccentricities remain close to the forced value, the resulting circular distribution is tightly clustered around it, and the deviation from the initial configuration is negligible. Consequently, particles that began in a narrow, eccentric configuration will remain in that state.
\end{enumerate}

The formation and long-term maintenance of a narrow, eccentric ring therefore requires that the grain distribution remains concentrated around the forced eccentricity in the $hk$-plane. To first order, the forced eccentricity can be set by the properties of the outermost planet. Using this approximation, we obtained forced components at the end of the simulation of $(e_{\rm forced},\varpi_{\rm forced})=(0.047,\ 225.29\degree)$, comparable to those of the largest grains, $(e_{\rm disk},\varpi_{\rm disk})=(0.066,\ 232.41\degree)$. If the trends observed in the simulations persist, these values may converge over longer evolutionary timescales, becoming the infall onto a PPD, a potential progenitor of the Fomalhaut-like debris disks.

\section{Conclusion}
\label{sec:conclusion}

We investigated the impact of environmental perturbations --- namely stellar flybys and gas infall events --- on the evolution of a protoplanetary disk hosting a resonant pair of planets. By performing three-dimensional hydrodynamical simulations, we characterized the morphological and dynamical signatures of a representative Class II protoplanetary disk undergoing interactions with its environment. We highlighted potential dust structures that could persist into the debris-disk phase as tracers of its dynamical history after gas dispersal. Our main findings are:
\begin{enumerate}
    \item 
    Both infall and flyby events significantly alter the global properties of the gaseous disk. Infall increases its mass budget, angular momentum, and energy, producing a dynamically excited disk. In contrast, stellar flybys truncate the disk, removing angular momentum and leading to a more compact disk configuration. 
    \item Dynamical excitation caused by these perturbations varies in both strength and duration. Infall results in a stronger and more prolonged excitation, leading to a more eccentric disk ($e\sim 0.04$--$0.08$; Figure~\ref{fig:disk_eccs}), as continuous material accretion extends its influence. Flybys, however, occur briefly, causing a quick but short-lived disturbance. These differences are evident in the evolution of the angular momentum deficit and the disk's eccentricity.
    \item The dust component responds differently from the gas, with a clear dependence on its degree of coupling to the gas. While strongly coupled grains closely follow the gas evolution, larger grains respond with a delay. Infall events create multiple eccentric dust populations, marking a clear distinction between coupled and decoupled dust grains, whereas flybys encourage radial migration and produce a more compact spatial distribution of dust. Importantly, an infall can lead to narrow, coherently eccentric dust disks that may persist into later evolutionary stages, potentially explaining Fomalhaut-like debris disk structures. 
    \item Planetary accretion is influenced by environmental interactions. Flybys enhance accretion onto both planets and promote the inward transport of solid material, potentially increasing planetary metallicity. Infall, in contrast, preferentially boosts accretion onto the inner planet, leading to a counterintuitive evolution of the accretion rates, in which the inner planet's accretion rate increases more than that of the outer planet, which is near the disk's inner edge.
\end{enumerate}

Our work compares two environmental effects that previous studies have examined separately. Additionally, it suggests a potential link between events during the protoplanetary disk phase and structures in the debris disk stage, as our results naturally explain the origin of narrow and eccentric debris disks. Further studies should confirm or rule out whether such an architecture remains \rev{after} long-term evolution. 

\begin{acknowledgements} 
This project has received funding from the European Research Council (ERC) under the European Union Horizon Europe programme (grant agreement No. 101042275, project Stellar-MADE). TDP is supported by a UKRI Stephen Hawking Fellowship and a Warwick Prize Fellowship, the latter made possible by a generous philanthropic donation. JC acknowledges support from the National Natural Science Foundation of China (NSFC) under grant number W2533008. DJP thanks IPAG, CNRS and UGA for hospitality and support towards local costs during his 2025 sabbatical in Grenoble where this work was completed. 
\end{acknowledgements}

\section*{Data availability}
The data required to reproduce all figures
in this paper are available at \href{10.5281/zenodo.20307800}{10.5281/zenodo.20307800}.

We utilized the following public software: {\sc Phantom}, \href{https://github.com/danieljprice/phantom}{https://github.com/danieljprice/phantom} \citep{PricePH+2018b}, and {\sc Splash}, \href{https://github.com/danieljprice/splash}{https://github.com/danieljprice/splash} \citep{Price2007}

\bibliographystyle{aa}
\bibliography{paper}

\appendix 

\section{Planetary orbital evolution}
\label{app:pl}

\begin{figure*}
\centering
\begin{center}
    \includegraphics[width=1\textwidth]{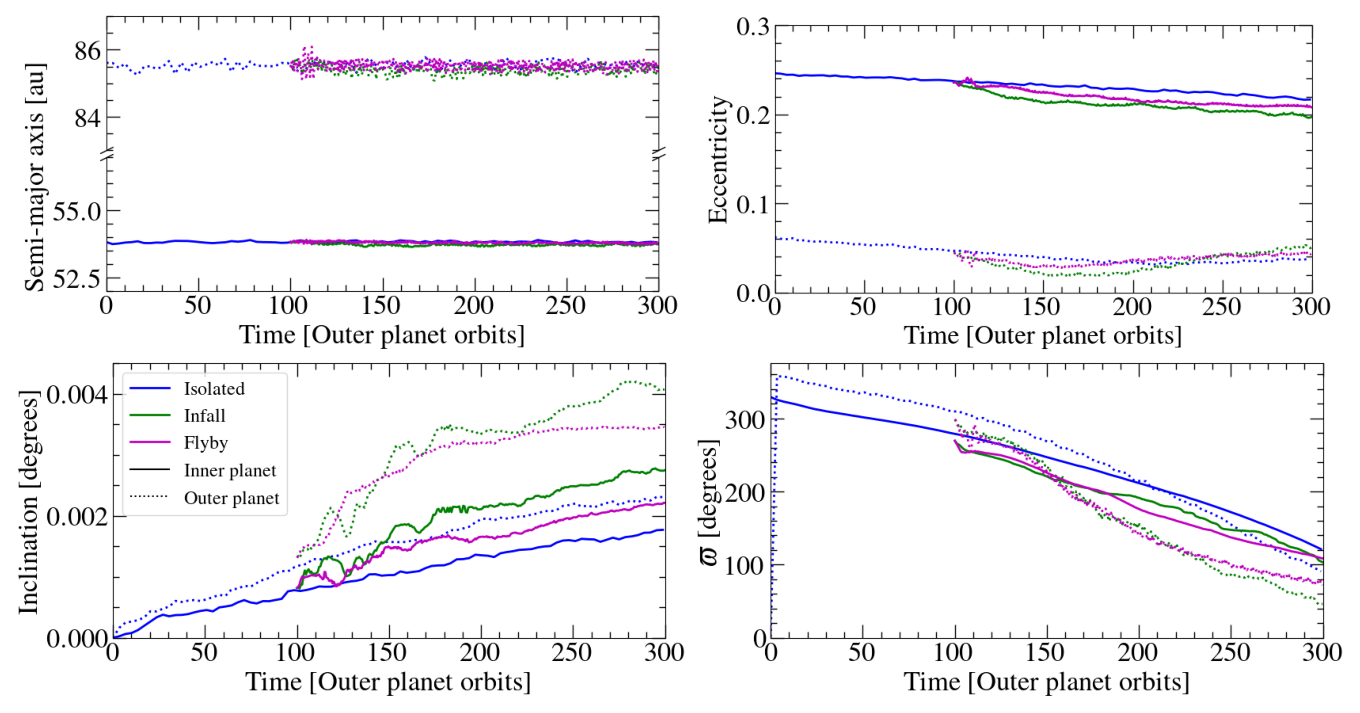} 
    \caption{Evolution of the semi-major axis (upper left), eccentricity (upper right), inclination (lower left), and longitude of pericentre (lower right) of the inner and outer planets, shown by solid and dotted lines, respectively. The color coding is the same as in Figs.~\ref{fig:Gas_profile}, \ref{fig:Gas_quantities}, \ref{fig:e_free}, and \ref{fig:Planet_mdot}.}
    \label{fig:pl_orbit}
\end{center}
\end{figure*}

\rev{Figure~\ref{fig:pl_orbit} shows the evolution of the semi-major axis, eccentricity, inclination, and longitude of pericentre for both planets. The semi-major axes remain essentially unchanged in all cases, indicating that neither planet undergoes significant migration. For the remaining orbital elements, the perturbed cases closely follow the evolution observed in the Isolated case. Although small variations are present, their amplitudes are negligible. Therefore, the main effect of the environmental perturbations is not a direct modification of the planetary orbits, but rather the alteration of the disk architecture and planetary accretion discussed in Sect.~\ref{sec:results}.}

\label{lastpage}
\end{document}